\documentclass{article}

\usepackage{arxiv}
\usepackage[utf8]{inputenc}
\usepackage[T1]{fontenc}
\usepackage[hidelinks]{hyperref}
\usepackage{booktabs}
\usepackage{nicefrac}
\usepackage{microtype}
\usepackage{natbib}
\usepackage{url}
\usepackage{enumitem}
\usepackage{amsmath,amssymb,amsfonts}
\usepackage{algorithmic}
\usepackage{graphicx}
\usepackage{textcomp}
\usepackage{xcolor}
\usepackage{xspace}
\usepackage{balance}
\usepackage{booktabs}
\usepackage{multirow}
\usepackage{authblk}
\usepackage[many]{tcolorbox}

\newcommand{\FOne}{$\text{F}_1$}

\title{Exploiting Token and Path-based Representations of Code for Identifying Security-Relevant Commits}

\author[1]{Achyudh Ram}
\author[1]{Ji Xin}
\author[1]{Meiyappan Nagappan}
\author[1]{Yaoliang Yu}
\author[2]{Roc\'io Cabrera Lozoya}
\author[2]{Antonino Sabetta}
\author[1]{Jimmy Lin}

\affil[1]{David R. Cheriton School of Computer Science, University of Waterloo}
\affil[2]{SAP Security Research}

\begin{document}
\maketitle

\begin{abstract}
Public vulnerability databases such as CVE and NVD account for only 60\% of security vulnerabilities present in open-source projects, and are known to suffer from inconsistent quality.
Over the last two years, there has been considerable growth in the number of known vulnerabilities across projects available in various repositories such as NPM and Maven Central.
Such an increasing risk calls for a mechanism to infer the presence of security threats in a timely manner.
We propose novel hierarchical deep learning models for the identification of security-relevant commits from either the commit diff or the source code for the Java classes.
By comparing the performance of our model against code2vec, a state-of-the-art model that learns from path-based representations of code, and a logistic regression baseline, we show that deep learning models show promising results in identifying security-related commits.
We also conduct a comparative analysis of how various deep learning models learn across different input representations and the effect of regularization on the generalization of our models.
\end{abstract}


\section{Introduction}

The use of open-source software has been steadily increasing for some time now, with the number of Java packages in Maven Central doubling in 2018.
However, \citet{snyk2019state} states that there has been an 88\% growth in the number of vulnerabilities reported over the last two years.
In order to develop secure software, it is essential to analyze and understand security vulnerabilities that occur in software systems and address them in a timely manner.
While there exist several approaches in the literature for identifying and managing security vulnerabilities, \citet{ponta2018beyond} show that an effective vulnerability management approach must be code-centric.
Rather than relying on metadata, efforts must be based on analyzing vulnerabilities and their fixes at the code level. 

Common Vulnerabilities and Exposures (CVE)\footnote{https://cve.mitre.org/} is a list of publicly known cybersecurity vulnerabilities, each with an identification number.
These entries are used in the National Vulnerability Database (NVD),\footnote{https://nvd.nist.gov/} the U.S. government repository of standards based vulnerability management data.
The NVD suffers from poor coverage, as it contains only 10\% of the open-source vulnerabilities that have received a CVE identifier \citep{sabetta2018practical}.
This could be due to the fact that a number of security vulnerabilities are discovered and fixed through informal communication between maintainers and their users in an issue tracker.
To make things worse, these public databases are too slow to add vulnerabilities as they lag behind a private database such as Snyk's DB by an average of 92 days \citep{snyk2019state}
All of the above pitfalls of public vulnerability management databases (such as NVD) call for a mechanism to automatically infer the presence of security threats in open-source projects, and their corresponding fixes, in a timely manner.

We propose a novel approach using deep learning in order to identify commits in open-source repositories that are security-relevant.
We build regularized hierarchical deep learning models that encode features first at the file level, and then aggregate these file-level representations to perform the final classification.
We also show that code2vec, a model that learns from path-based representations of code and claimed by \citet{alon2018general} to be suitable for a wide range of source code classification tasks, performs worse than our logistic regression baseline.

In this study, we seek to answer the following research questions:

\begin{itemize}[leftmargin=*]
    \item \textbf{RQ1:} \textit{Can we effectively identify security-relevant commits using only the commit diff?}
    For this research question, we do not use any of the commit metadata such as the commit message or information about the author. 
    We treat source code changes like unstructured text without using path-based representations from the abstract syntax tree.
    \item \textbf{RQ2:} \textit{Does extracting class-level features before and after the change instead of using only the commit diff improve the identification of security-relevant commits?} 
    For this research question, we test the hypothesis that the source code of the entire Java class contains more information than just the commit diff and could potentially improve the performance of our model.
    \item \textbf{RQ3:} \textit{Does exploiting path-based representations of Java source code before and after the change improve the identification of security-relevant commits?}
    For this research question, we test whether code2vec, a state-of-the-art model that learns from path-based representations of code, performs better than our model that treats source code as unstructured text.
    \item \textbf{RQ4:} \textit{Is mining commits using regular expression matching of commit messages an effective means of data augmentation for improving the identification of security-relevant commits?}
    Since labelling commits manually is an expensive task, it is not easy to build a dataset large enough to train deep learning models.
    For this research question, we explore if collecting coarse data samples using a high-precision approach is an effective way to augment the ground-truth dataset.  
\end{itemize}

The main contributions of this paper are:
\begin{itemize}[leftmargin=*]
	\item Novel hierarchical deep learning models for the identification of security-relevant commits based on either the diff or the modified source code of the Java classes.
	\item A comparative analysis of how various deep learning models perform across different input representations and how various regularization techniques help with the generalization of our models. 
\end{itemize}

We envision that this work would ultimately allow for monitoring open-source repositories in real-time, in order to automatically detect security-relevant changes such as vulnerability fixes. 

\section{Background and Related Work}

\subsection{Neural Networks for Text Classification}

In computational linguistics, there has been a lot of effort over the last few years to create a continuous higher dimensional vector space representation of words, sentences, and even documents such that similar entities are closer to each other in that space \citep{mikolov2013efficient, le2014distributed, lebret2015sum}.
\citet{mikolov2013efficient} introduced word2vec, a class of two-layer neural network models that are trained on a large corpus of text to produce word embeddings for natural language. 
Such learned distributed representations of words have accelerated the application of deep learning techniques for natural language processing (NLP) tasks \citep{bengio2003neural}. 

\citet{kim2014convolutional} show that convolutional neural networks (CNNs) can achieve state-of-the-art results in single-sentence sentiment prediction, among other sentence classification tasks. 
In this approach, the vector representations of the words in a sentence are concatenated vertically to create a two-dimensional matrix for each sentence. 
The resulting matrix is passed through a CNN to extract higher-level features for performing the classification. 
\citet{yang2016hierarchical} introduce the hierarchical attention network (HAN), where a document vector is progressively built by aggregating important words into sentence vectors, and then aggregating important sentences vectors into document vectors.

Deep neural networks are prone to overfitting due to the possibility of the network learning complicated relationships that exist in the training set but not in unseen test data. 
Dropout prevents complex co-adaptations of hidden units on training data by randomly removing (i.e. dropping out) hidden units along with their connections during training \citep{srivastava2014dropout}. 
Embedding dropout, used by \citet{merity2018regularizing} for neural language modeling, performs dropout on entire word embeddings. 
This effectively removes a proportion of the input tokens randomly at each training iteration, in order to condition the model to be robust against missing input.

While dropout works well for regularizing fully-connected layers, it is less effective for convolutional layers due to the spatial correlation of activation units in convolutional layers. 
There have been a number of attempts to extend dropout to convolutional neural networks \citep{wu2015towards}. 
DropBlock is a form of structured dropout for convolutional layers where units in a contiguous region of a feature map are dropped together \citep{ghiasi2018dropblock}.

\subsection{Learning Embeddings for Source Code}

While building usable embeddings for source code that capture the complex characteristics involving both syntax and semantics is a challenging task, such embeddings have direct downstream applications in tasks such as semantic code clone detection, code captioning, and code completion \citep{alon2018code2seq, yahav2018programs}. 
In the same vein as \citet{mikolov2013efficient}, neural networks have been used for representing snippets of code as continuous distributed vectors \citep{alon2018code2vec}. 
They represent a code snippet as a bag of contexts and each context is represented by a context vector, followed by a path-attention network that learns how to aggregate these context vectors in a weighted manner. 

A number of other code embedding techniques are also available in the literature.
\citet{henkel2018code} learn word embeddings from abstractions of traces obtained from the symbolic execution of a program. 
They evaluate their learned embeddings on a benchmark of API-usage analogies extracted from the Linux kernel and achieved 93\% top-1 accuracy. 
\citet{hamelsmu2018semantic} describe a pipeline that leverages deep learning for semantic search of code. 
To achieve this, they train a sequence-to-sequence model that learns to summarize Python code by predicting the corresponding docstring from the code blob, and in the process provide code representations for Python. 

\subsection{Identifying Security Vulnerabilities}

There exist a handful of papers in software engineering that perform commit classification to identify security vulnerabilities or fixes. 
\citet{zhou2017automated} describe an efficient vulnerability identification system geared towards tracking large-scale projects in real time using latent information underlying commit messages and bug reports in open-source projects. 
While \citet{zhou2017automated} classify commits based on the commit message, we use only the commit diff or the corresponding source code as features for our model. 
\citet{sabetta2018practical} propose a machine learning approach to identify security-relevant commits. 
However, they treat source code as documents written in natural language and use well-known document classification methods to perform the actual classification. 
\citet{bosu2014identifying} conduct an analysis to identify which security vulnerabilities can be discovered during code review, or what characteristics of developers are likely to introduce vulnerabilities. 

\section{Experimental Setup}

This section details the methodology used in this study to build the training dataset, the models used for classification and the evaluation procedure. 
All of the experiments are conducted on Python 3.7 running on an Intel Core i7 6800K CPU and a Nvidia GTX 1080 GPU. 
All the deep learning models are implemented in PyTorch 0.4.1~\citep{paszke2017automatic}, while Scikit-learn 0.19.2~\citep{scikit-learn} is used for computing the tf--idf vectors and performing logistic regression.

For training our classification models, we use a manually-curated dataset of publicly disclosed vulnerabilities in 205 distinct open-source Java projects mapped to commits fixing them, provided by \citet{ponta2019manually}. 
These repositories are split into training, validation, and test splits containing 808, 265, and 264 commits, respectively. 
In order to minimize the occurrence of duplicate commits in two of these splits (such as in both training and test), commits from no repository belong to more than one split. 
However, 808 commits may not be sufficient to train deep learning models. 
Hence, in order to answer RQ4, we augment the training split with commits mined using regular expression matching on the commit messages from the same set of open-source Java projects. 
This almost doubles the number of commits in the training split to 1493. 
We then repeat our experiments for the first three research questions on the augmented dataset, and evaluate our trained models on the same validation and test splits. 

We also compare the quality of randomly-initialized embeddings with pre-trained ones. 
Since the word2vec embeddings only need unlabelled data to train, the data collection and preprocessing stage is straightforward. 
GitHub, being a very large host of source code, contains enough code for training such models. 
However, a significant proportion of code in GitHub does not belong to engineered software projects \citep{munaiah2017curating}. 
To reduce the amount of noise in our training data, we filter repositories based on their size, commit history, number of issues, pull requests, and contributors, and build a corpus of the top 1000 Java repositories. 
We limit the number of repositories to 1000 due to GitHub API limitations. 
It is worth noting that using a larger training corpus might provide better results. 
For instance, code2vec is pre-trained on a corpus that is ten times larger. 

\begin{figure}[t]
    \centering
    \includegraphics[width=0.5\linewidth]{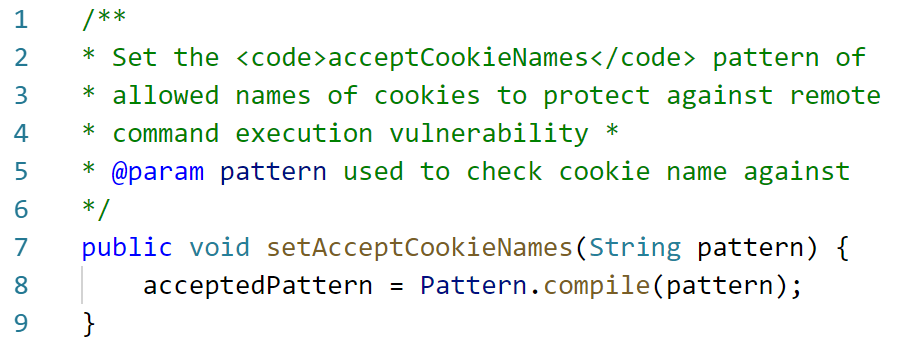}
    \caption{A code snippet from Apache Struts with the Javadoc stating which vulnerability was addressed}
    \label{fig:snippet}
\end{figure}

To extract token-level features for our model, we use the lexer and tokenizer provided as a part of the Python javalang library.\footnote{https://pypi.org/project/javalang/} 
We ensure that we only use the code and not code comments or metadata, as it is possible for comments or commit messages to include which vulnerabilities are fixed, as shown in Figure~\ref{fig:snippet}. 
Our models would then overfit on these features rather than learning the features from the code. 
For extracting path-based representations from Java code, we use ASTMiner.\footnote{https://github.com/vovak/astminer}

\section{Model}

\subsection{Training Word2vec Embeddings}

We learn token-level vectors for code using the CBOW architecture \citep{mikolov2013efficient}, with negative sampling and a context window size of 5. 
Using CBOW over skip-gram is a deliberate design decision. While skip-gram is better for infrequent words, we felt that it is more important to focus on the more frequent words (inevitably, the keywords in a programming language) when it comes to code. 
Since we only perform minimal preprocessing on the code (detailed below), the most infrequent words will usually be variable identifiers. 
Following the same line of reasoning, we choose negative sampling over hierarchical-softmax as the training algorithm. 

We do not normalize variable identifiers into generic tokens as they could contain contextual information. 
However, we do perform minimal preprocessing on the code before training the model. This includes:

\begin{enumerate}[leftmargin=*]
    \item The removal of comments and whitespace when performing tokenization using a lexer.
    \item The conversion of all numbers such as integers and floating point units into reserved tokens.
    \item The removal of tokens whose length is greater than or equal to 64 characters.
    \item Thresholding the size of the vocabulary to remove infrequent tokens.
\end{enumerate}

\begin{figure*}[t]
    \centering
    \includegraphics[width=0.95\linewidth]{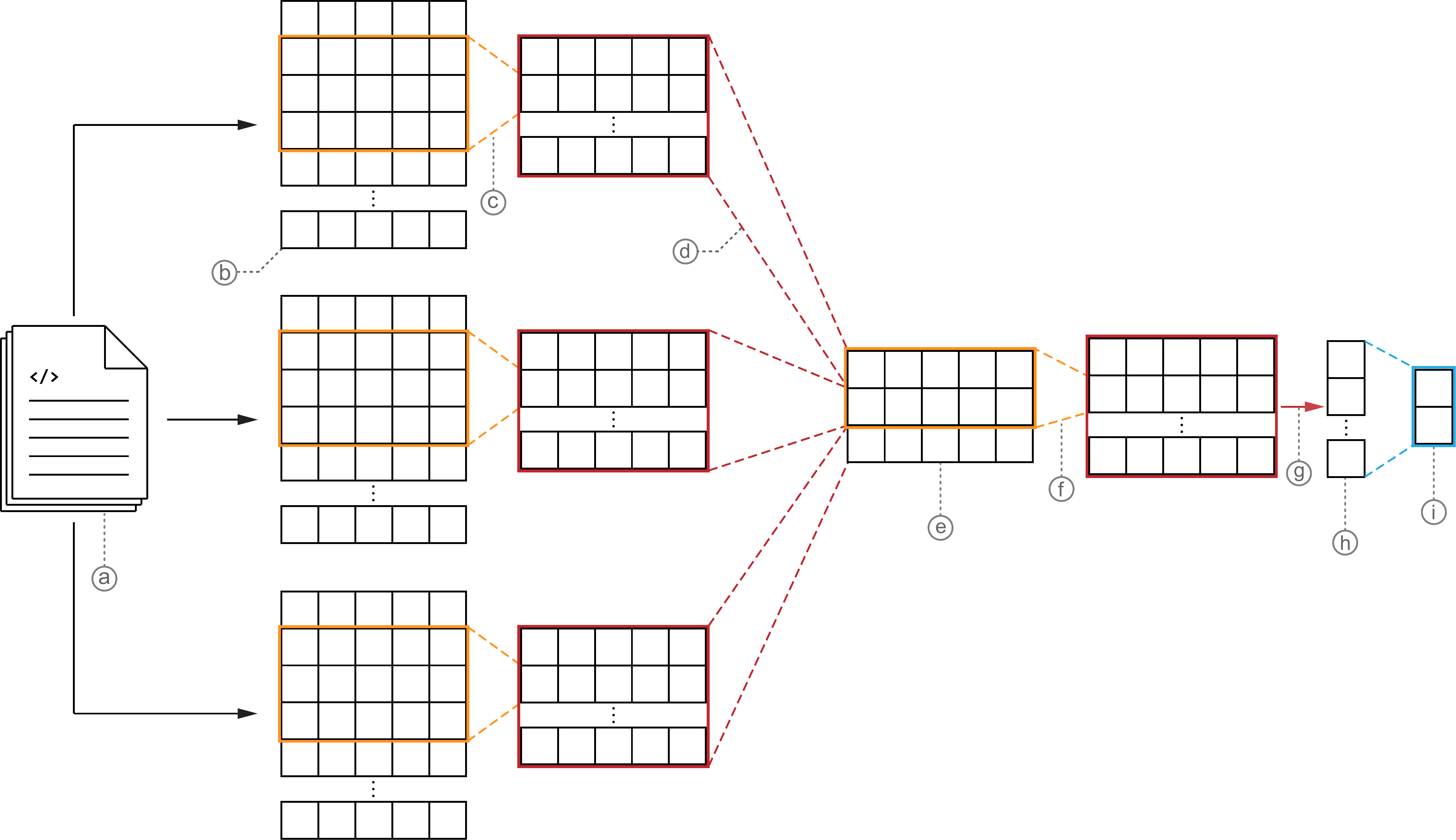}
    \caption{Illustration of our H-CNN model for learning on diff tokens, where the labels are the following:\ (a) source code diffs from multiple files, (b) stacked token embeddings, (c) convolutional feature extraction, (d) max-pool across time, (e) file-level feature maps, (f) convolutional feature extraction, (g) max-pool across time, (h) commit-level feature vector, and (i) softmax output.}
    \label{fig:hybrid_cnn}
\end{figure*}

\subsection{Identifying Security Vulnerabilities}

We modify our model accordingly for every research question, based on changes in the input representation. 
To benchmark the performance of our deep learning models, we compare them against a logistic regression (LR) baseline that learns on one-hot representations of the Java tokens extracted from the commit diffs. 
For all of our models, we employ dropout on the fully-connected layer for regularization. 
We use Adam \citep{kingma2014adam} for optimization, with a learning rate of 0.001, and batch size of 16 for randomly initialized embeddings and 8 for pre-trained embeddings.

For RQ1, we use a hierarchical CNN (H-CNN) with either randomly-initialized or pre-trained word embeddings in order to extract features from the commit diff. 
We represent the commit diff as a concatenation of 300-dimensional vectors for each corresponding token from that diff. 
This resultant matrix is then passed through three temporal convolutional layers in parallel, with filter windows of size 3, 5, and 7. 
A temporal max-pooling operation is applied to these feature maps to retain the feature with the highest value in every map. 
We also present a regularized version of this model (henceforth referred to as HR-CNN) with embedding dropout applied on the inputs, and DropBlock on the activations of the convolutional layers.

For RQ2, we made a modification to both the H-CNN and HR-CNN models in order to extract features from the source code for the Java classes before and after the commit. 
Both of these models use a siamese architecture between the two CNN-based encoders as shown in Figure \ref{fig:hybrid_cnn}. 
We then concatenate the results from both of these encoders and pass it through a fully-connected layer followed by softmax for prediction.

For RQ3, we adapt the code2vec model used by \citet{alon2018code2vec} for predicting method names into a model for predicting whether a commit is security-relevant by modifying the final layer. 
We then repeat our experiments on both the ground-truth and augmented dataset. 

\section{Results and Discussion}

The results for all of our models on both the ground-truth and augmented datasets are given in Table \ref{tab:results}.

\begin{table*}[t]
\centering
\scalebox{0.95}{
\begin{tabular}{@{}llllllllllll@{}}
\toprule
\multirow{2}{*}{\textbf{\#}} & \multirow{2}{*}{\textbf{Input features}} & \multirow{2}{*}{\textbf{Model}} & \multirow{2}{*}{\textbf{Embedding}} & \multicolumn{4}{c}{\textbf{Validation}} & \multicolumn{4}{c}{\textbf{Test}} \\ \cmidrule(lr){5-8} \cmidrule(lr){9-12} 
 &  &  &  & Acc. & P & R & \FOne & Acc. & P & R & \FOne \\ \midrule
\multicolumn{12}{c}{\textit{Ground-truth dataset}} \\ \midrule
1 & Diff Tokens & LR & One-hot & 0.644 & 0.648 & 0.913 & 0.758 & 0.630 & 0.645 & 0.877 & 0.743 \\
2 &  & H-CNN & Random & 0.636 & 0.635 & 0.950 & 0.761 & 0.657 & 0.645 & 0.975 & 0.776 \\
3 &  &  & Pre-trained & 0.682 & 0.702 & 0.832 & 0.761 & 0.600 & 0.649 & 0.753 & 0.697 \\
4 &  & HR-CNN & Random & 0.674 & 0.676 & 0.894 & 0.770 & 0.645 & 0.660 & 0.864 & 0.749 \\
5 &  &  & Pre-trained & 0.633 & 0.629 & 0.969 & 0.763 & 0.653 & 0.641 & \textbf{0.981} & 0.776 \\
6 & Paired-code Tokens & H-CNN & Random & 0.633 & 0.633 & 0.944 & 0.758 & 0.649 & 0.643 & 0.957 & 0.769 \\
7 &  &  & Pre-trained & 0.663 & 0.651 & 0.963 & 0.777 & 0.630 & 0.632 & 0.944 & 0.757 \\
8 &  & HR-CNN & Random & 0.674 & 0.671 & 0.913 & 0.774 & 0.668 & 0.673 & 0.889 & 0.766 \\
9 &  &  & Pre-trained & \textbf{0.746} & \textbf{0.761} & 0.851 & \textbf{0.804} & \textbf{0.725} & \textbf{0.726} & 0.883 & \textbf{0.797} \\
10 & Paired-AST Paths & Code2Vec & Random & 0.622 & 0.619 & \textbf{1.000} & 0.764 & 0.613 & 0.612 & 0.987 & 0.756 \\ \midrule
\multicolumn{12}{c}{\textit{Augmented dataset}} \\ \midrule
11 & Diff Tokens & LR & One-hot & 0.697 & 0.731 & 0.795 & 0.762 & 0.653 & 0.716 & 0.716 & 0.716 \\
12 &  & H-CNN & Random & 0.663 & 0.658 & 0.932 & 0.771 & 0.608 & 0.622 & 0.914 & 0.740 \\
13 &  &  & Pre-trained & 0.659 & 0.732 & 0.696 & 0.713 & 0.623 & \textbf{0.704} & 0.660 & 0.682 \\
14 &  & HR-CNN & Random & 0.663 & 0.658 & 0.932 & 0.771 & 0.608 & 0.622 & 0.914 & 0.740 \\
15 &  &  & Pre-trained & 0.648 & 0.739 & 0.652 & 0.693 & 0.596 & 0.692 & 0.611 & 0.649 \\
16 & Paired-code Tokens & H-CNN & Random & 0.610 & 0.610 & \textbf{1.000} & 0.758 & 0.611 & 0.611 & \textbf{1.000} & 0.759 \\
17 &  &  & Pre-trained & 0.610 & 0.610 & \textbf{1.000} & 0.758 & 0.623 & 0.618 & \textbf{1.000} & \textbf{0.764} \\
18 &  & HR-CNN & Random & 0.610 & 0.610 & \textbf{1.000} & 0.758 & 0.611 & 0.611 & \textbf{1.000} & 0.759 \\
19 &  &  & Pre-trained & \textbf{0.742} & \textbf{0.736} & 0.901 & \textbf{0.810} & \textbf{0.668} & 0.683 & 0.852 & 0.758 \\
20 & Paired-AST Paths & Code2Vec & Random & 0.629 & 0.624 & 1.000 & 0.768 & 0.625 & 0.621 & 0.974 & 0.759 \\ \bottomrule
\end{tabular}}
\caption{Results for each model on the validation and test splits; best values are bolded.}
\label{tab:results}
\end{table*}

\textbf{RQ1:} \textit{Can we effectively identify security-relevant commits using only the commit diff?} 

Without using any of the metadata present in a commit, such as the commit message or information about the author, we are able to correctly classify commits based on their security-relevance with an accuracy of 65.3\% and \FOne of 77.6\% on unseen test data. 
Table \ref{tab:results}, row~5, shows that using our regularized HR-CNN model with pre-trained embeddings provides the best overall results on the test split when input features are extracted from the commit diff. 
Table \ref{tab:results}, row~3, shows that while H-CNN provides the most accurate results on the validation split, it doesn't generalize as well to unseen test data. 
While these results are usable, H-CNN and HR-CNN only perform 3 points better than the LR baseline (Table \ref{tab:results}, row~1) in terms of \FOne and 2 points better in terms of accuracy. 

\textbf{RQ2:} \textit{Does extracting class-level features before and after the change instead of using only the commit diff improve the identification of security-relevant commits?} 

When extracting features from the complete source code of the Java classes which are modified in the commit, the performance of HR-CNN increases noticeably. 
Table \ref{tab:results}, row~9, shows that the accuracy of HR-CNN when using pre-trained embeddings increases to 72.6\% and \FOne increases to 79.7\%. 
This is considerably above the LR baseline and justifies the use of a more complex deep learning model. 
Meanwhile, the performance of H-CNN with randomly-initialized embeddings (Table \ref{tab:results}, row~6) does not improve when learning on entire Java classes, but there is a marked improvement in \FOne of about 6 points when using pre-trained embeddings. 
Hence, we find that extracting class-level features from the source code before and after the change, instead of using only the commit diff, improves the identification of security-relevant commits.

\textbf{RQ3:} \textit{Does exploiting path-based representations of the Java classes before and after the change improve the identification of security-relevant commits?} 

Table \ref{tab:results}, row~10, shows that training the modified code2vec model to identify security-aware commits from scratch results in a model that performs worse than the LR baseline. 
The model only achieves an accuracy of 63.8\% on the test split, with an \FOne score of 72.7\%, which is two points less than that of LR. 
The code2vec model performs much worse compared to H-CNN and HR-CNN with randomly-initialized embeddings. 
Hence, learning from a path-based representation of the Java classes before and after the change does not improve the identification of security-relevant commits---at least with the code2vec approach.

\textbf{RQ4:} \textit{Is mining commits using regular expression matching of commit messages an effective means of data augmentation for improving the identification of security-relevant commits?} 

The results in Table \ref{tab:results}, rows 11 to 20, show that collecting coarse data samples using regular expression matching for augmenting the ground-truth training set is not effective in increasing the performance of our models. 
This could possibly be due to the coarse data samples being too noisy or the distribution of security-relevant commits in the coarse dataset not matching that of the unseen dataset. 
The latter might have been due to the high-precision mining technique used, capturing only a small subset of security vulnerabilities. 

\subsection{Threats to Validity}

The lexer and tokenizer we use from the \texttt{javalang} library target Java 8. 
We are not able to verify that all the projects and their forks in this study are using the same version of Java. 
However, we do not expect considerable differences in syntax between Java 7 and Java 8 except for the introduction of lambda expressions.\footnote{https://www.jcp.org/en/jsr/detail?id=335}

There is also a question of to what extent the 635 publicly disclosed vulnerabilities used for evaluation in this study represent the vulnerabilities found in real-world scenarios. 
While creating larger ground-truth datasets would always be helpful, it might not always be possible. 
To reduce the possibility of bias in our results, we ensure that we don't train commits from the same projects that we evaluate our models on. 
We also discard any commits belonging to the set of evaluation projects that are mined using regular expression matching.

We directly train code2vec on our dataset without pre-training it, in order to assess how well path-based representations perform for learning on code, as opposed to token-level representations on which H-CNN and HR-CNN are based. 
However, \citet{alon2018code2vec} pre-trained their model on 10M Java classes. 
It is possible that the performance of code2vec is considerably better than the results in Table \ref{tab:results} after pre-training.
Furthermore, our findings apply only to this particular technique to capturing path-based representations, not the approach in general.
However, we leave both issues for future work. 

\section{Conclusions and Future Work}

In this study, we propose a novel hierarchical deep learning model for the identification of security-relevant commits and show that deep learning has much to offer when it comes to commit classification.
We also make a case for pre-training word embeddings on tokens extracted from Java code, which leads to performance improvements.
We are able to further improve the results using a siamese architecture connecting two CNN-based encoders to represent the modified files before and after a commit.

Network architectures that are effective on a certain task, such as predicting method names, are not necessarily effective on related tasks. 
Thus, choices between neural models should be made considering the nature of the task and the amount of training data available. 
Based on the model's ability to predict method names in files across different projects, \citet{alon2018code2vec} claim that code2vec can be used for a wide range of programming language processing tasks. 
However, for predicting the security relevance of commits, H-CNN and HR-CNN appear to be much better than code2vec.

A potential research direction would be to build language models for programming languages based on deep language representation models. 
Neural networks are becoming increasingly deeper and complex in the NLP literature, with significant interest in deep language representation models such as ELMo, GPT, and BERT \citep{peters2018elmo, radford2018gpt, devlin2018bert}. \citet{devlin2018bert} show strong empirical performance on a broad range of NLP tasks. 
Since all of these models are pre-trained in an unsupervised manner, it would be easy to pre-train such models on the vast amount of data available on GitHub. 

Deep learning models are known for scaling well with more data. 
However, with less than 1,000 ground-truth training samples and around 1,800 augmented training samples, we are unable to exploit the full potential of deep learning. 
A reflection on the current state of labelled datasets in software engineering (or the lack thereof) throws light on limited practicality of deep learning models for certain software engineering tasks \citep{lin2018sentiment}. 
As stated by \citet{allamanis2018survey}, just as research in NLP changed focus from brittle rule-based expert systems to statistical methods, software engineering research should augment traditional methods that consider only the formal structure of programs with information about the statistical properties of code. 
Ongoing research on pre-trained code embeddings that don't require a labelled dataset for training is a step in the right direction. 
Drawing parallels with the recent history of NLP research, we are hoping that further study in the domain of code embeddings will considerably accelerate progress in tackling software problems with deep learning.

\section*{Acknowledgments}
We would like to thank SAP and NSERC for their support towards this project.

\bibliographystyle{plainnat}
\bibliography{main}

\end{document}